\documentclass[reprint,prb,amsmath,amssymb,superscriptaddress]{revtex4-2}
\usepackage{graphicx} 
\usepackage{amsfonts, amsmath, amsthm, amssymb}
\usepackage{graphicx}
\usepackage{dcolumn}
\usepackage{multirow} 
\usepackage{bm}
\usepackage{xcolor}
\usepackage{float}
\usepackage{caption}
\usepackage{subcaption}
\usepackage[normalem]{ulem} 
\usepackage[bottom]{footmisc}
\usepackage{footnote}
\usepackage[version=4,arrows=pgf-filled
]{mhchem}

\usepackage{amsmath} 

\newcommand{\exciting}{{\usefont{T1}{lmtt}{}{n}exciting}}

\newcommand{\ie}{{\it i.e.}, }
\newcommand{\eg}{{\it e.g.}, }

\begin{document}

\title{Precision benchmarks for solids: \textit{${G_0W_0}$} calculations with different basis sets}
\author{Maryam Azizi}
\email{maryam.azizi@uclouvain.be}
\affiliation{Université\ Catholique\ de\ Louvain,\ Louvain-la-Neuve, Belgium}
\affiliation{European Theoretical Spectroscopic Facility (ETSF)}
\author{Francisco A. Delesma}
\affiliation{Department of Applied Physics, Aalto University, FI-02150 Espoo, Finland}
\affiliation{Faculty for Chemistry and Food Chemistry, Technische Universit\"at Dresden, 01062 Dresden, Germany}
\author{Matteo Giantomassi}
\affiliation{Université\ Catholique\ de\ Louvain,\ Louvain-la-Neuve, Belgium}
\affiliation{European Theoretical Spectroscopic Facility (ETSF)}
\author{Davis Zavickis}
\affiliation{University\ of\ Latvia,\ Riga,\ Latvia}
\author{Mikael Kuisma}
\affiliation{Danish\ Technical\ University,\ Lyngby,\ Denmark}
\author{Kristian Thyghesen}
\affiliation{Danish\ Technical\ University,\ Lyngby,\ Denmark}
\author{Dorothea Golze}
\affiliation{Faculty for Chemistry and Food Chemistry, Technische Universit\"at Dresden, 01062 Dresden, Germany}
\author{Alexander Buccheri}
\affiliation{Humboldt-Universit\"at  zu Berlin, Berlin, Germany}
\altaffiliation[New address of A. Buccheri: ]{Max Planck Institute for the Structure and Dynamics of Matter, Hamburg, Germany}
\author{Min-Ye Zhang}
\affiliation{The NOMAD Laboratory at the Fritz Haber Institute of the Max Planck Society, Berlin, Germany}
\author{Patrick Rinke}
\affiliation{Department of Applied Physics, Aalto University, FI-02150 Espoo, Finland}
\affiliation{Physics Department, TUM School of Natural Sciences, Technical University of Munich, Garching, Germany}
\affiliation{Atomistic Modelling Center, Munich Data Science Institute, Technical University of Munich, Garching, Germany}
\author{Claudia Draxl}
\affiliation{Humboldt-Universit\"at  zu Berlin, Berlin, Germany}
\affiliation{European Theoretical Spectroscopic Facility (ETSF)}
\author{Andris Gulans}
\affiliation{University\ of\ Latvia,\ Riga,\ Latvia}
\author{Xavier Gonze}
\affiliation{Université\ Catholique\ de\ Louvain,\ Louvain-la-Neuve, Belgium}
\affiliation{European Theoretical Spectroscopic Facility (ETSF)}


\date{\today}
\begin{abstract} 
The $GW$ approximation within many-body perturbation theory is the state of the art for computing quasiparticle energies in solids. Typically, Kohn-Sham (KS) eigenvalues and eigenfunctions, obtained from a Density Functional Theory (DFT) calculation are used as a starting point to build the Green’s function $G$ and the screened Coulomb interaction $W$, yielding the one-shot $G_0W_0$ self-energy if no further update of these quantities are made. Multiple implementations exist for both the DFT and the subsequent $G_0W_0$ calculation, leading to possible differences in quasiparticle energies. In the present work, the $G_0W_0$ quasiparticle energies for states close to the band gap are calculated for six crystalline solids, using four different codes: Abinit, \exciting{}, FHI-aims, and GPAW. This comparison helps to assess the impact of basis-set types (planewaves versus localized orbitals) and the treatment of core and valence electrons (all-electron full potentials versus pseudopotentials). The impact of unoccupied states as well as the algorithms for solving the quasiparticle equation are also briefly discussed. For the KS-DFT band gaps, we observe good agreement between all codes, with differences not exceeding 0.1 eV, while the $G_0W_0$ results deviate on the order of 0.1-0.3 eV. Between all-electron codes (FHI-aims and \exciting), the agreement is better than 15 meV for KS-DFT and, with one exception, about 0.1 eV for $G_0W_0$ band gaps. 
\end{abstract}
\maketitle

\section{\label{sec:introduction}Introduction}
The $GW$ approach~\cite{HedinGW} of many-body perturbation theory (MBPT) is the first-principles method of choice for the calculation of band structures of solids, and is becoming increasingly popular in molecular science too \cite{reviewgw2019}. For molecular systems, the $GW$ precision is well under control, as demonstrated by the $GW$100 benchmark efforts using multiple $GW$ implementations~\cite{van2015gw,Caruso2016,Maggio2017,Govoni2018}. However, benchmarking $GW$ calculations for solids is more challenging, not only due to the unfavorable scaling with system size but also due to the treatment of the long-range Coulomb interaction and the larger number of convergence parameters. As a consequence, systematic $GW$ benchmarks for materials across multiple codes are scarce, examples being Refs.~\cite{rangel2020reproducibility} and~\cite{Ren/etal:2021}. 

The original formulation of the $GW$ approximation foresees self-consistent updates of the Green's function, the dielectric screening, and the self-energy~\cite{HedinGW}. The usual practice, however, reduces the method to compute $G$ and $W$ from preliminary mean-field (DFT or Hartree-Fock) eigenenergies and eigenvalues, and combine them to get the self-energy without further updates~\cite{Onida2002,Setten2017,rangel2020reproducibility,Ren/etal:2021}. 
This approach is commonly known as $G_0W_0$. As a matter of fact, $G_0W_0$ offers competitive accuracy with a significantly reduced complexity and computational effort compared to, \eg fully self-consistent $GW$ or partially self-consistent $GW_0$. 

The accuracy (agreement with experimental data) obtained thanks to different electronic-structure methodologies has been increasingly studied during the past decades, along with improvements in precision, \eg improved numerical approximations and convergence, and better implementations of the same methodology. For an assessment of the methodologies used in the community, their accuracy must be considered for many basic properties of materials, like total energies, lattice parameters, not just the band gaps. This implies, however, an in-depth control of the precision (implementation and numerical parameters). The ``$\Delta$-factor" project has played an important role in the community. In this project, the DFT total energies obtained for 71 elemental solids using 40 different implementations (different codes, different basis sets, etc.) have been compared~\cite{Lejaeghere2016}. Such effort, focused on total energies, lattice parameters and bulk modulus, has been pursued over the years,
with improved coverage of materials~\cite{Bosoni2023}.

Considering the electronic band gap, Borlido and co-workers compiled a large data set of 472 materials designed for the efficient benchmarking of DFT exchange-correlation functionals~\cite{borlido2019large,borlido2020exchange}. From the comparison of experimental and theoretical band gaps, they assessed that the best functionals yield a mean absolute error of approximately 0.5 eV. Benchmarks on electronic band gaps at the $GW$ levels have been conducted on much smaller sets, comprising less than 30 materials \cite{shishkin2007self,Nabok2016, jiang2016gw, van2006quasiparticle, choi2016first, jiang2018revisiting, Grumet2018, Ren/etal:2021}. An exception is the study by van Setten \textit{et al.}~\cite{Setten2017}, which assessed a set of about 80 solids. Including partial self-consistency, an agreement of 0.15~eV with respect to experiment was reported for a benchmark set of 26 semiconductors and insulators, using an augmented planewave basis sets extended with local orbitals~\cite{jiang2016gw}. However, due to the slow convergence of $GW$, particularly with respect to basis set size~\cite{reviewgw2019}, precision benchmark studies remain necessary for a definitive assessment.

From a pragmatic perspective, the precision requirements for one methodology implemented in one code must be much smaller than the experimental uncertainty, \textit{i.e.}, of the order of tenths of an electronvolt \cite{borlido2019large}, and, similarly, must be much smaller than the characteristic difference between several first-principles methodologies and implementations. 
A discussion on the predictive power of the DFT or $GW$ methods, relative to experimental references,  is possible only if such precision is attained. In the DFT studies by Borlido and co-workers~\cite{borlido2019large,borlido2020exchange}, which employed various exchange-correlation functionals, only a single implementation was tested, raising questions about the numerical precision of the calculated band gaps. 
A critical aspect worth considering is how these results depend on factors such as the choice of the basis set and whether an all-electron or pseudopotential treatment is employed. These dependencies can manifest themselves not only at the DFT level, but are potentially amplified in $GW$ due to the larger number of convergence parameters. Establishing confidence in the precision of $GW$ calculations is not only necessary to benchmark existing conventional implementations but also crucial for the numerical validation of novel low-scaling $GW$ algorithms, which are currently emerging rapidly~\cite{liu2016,vlcek2017stochastic,Vlcek2018,Gao2018,wilhelm2018,foerster2020,Kim2020,Kutepov2020,duchemin2021,wilhelm2021,Gao2022,panades2023,graml2023lowscaling,Yeh2024,Shi2024,Azizi2024}.

Systematic numerical validation studies for solids comparing different codes are rare. Ren \textit{et al.}~\cite{Ren/etal:2021} restricted the discussion to $G_0W_0$ results from two all-electron codes, finding mean absolute deviations in the range of 0.15~eV for a set of 20 materials. Rangel and co-workers~\cite{rangel2020reproducibility} compared the results from three pseudopotential $G_0W_0$ implementations for a set of four materials: Si-diamond, \ce{TiO_2}-rutile, \ce{ZnO}-wurtzite and fcc gold. With judicious choices of approximations and using the same pseudopotentials, the converged $G_0W_0$ quasi-particle energies calculated with the different codes agree within 0.1 eV, addressing long-standing controversies surrounding the $G_0W_0$ results for difficult systems such as \ce{ZnO} and rutile.

In this study, we present $G_0W_0$ calculations for a set of six materials, including Si-diamond, \ce{TiO2}-rutile and \ce{ZnO}-wurtzite. In contrast to Refs.~\cite{Ren/etal:2021,rangel2020reproducibility}, we use four different codes implementing different approaches. (i) Abinit~\cite{Gonze2016,Gonze2020,Romero2020} is based on planewaves and norm-conserving pseudopotentials; (ii) the all-electron code \exciting~\cite{gulans2014exciting} implements different flavors of the linearized   augmented planewave plus local orbitals (LAPW+lo) method; (iii) the all-electron FHI-aims~\cite{FHIaims} employs numeric atom-centered orbitals; and (iv) GPAW~\cite{mortensen2024gpaw} uses planewaves with projector-augmented waves. 
We detail the parameters required for each of these codes for obtaining reproducible high-quality DFT and $G_0W_0$ data. While differences of less than or about 0.1 eV can be attained for DFT results, the $G_0W_0$ results differ by about 0.1-0.3 eV. The discrepancies typically reduce to within 0.1~eV when considering only the all-electron codes \exciting{} and FHI-aims.

The article is structured as follows: first, a brief summary of the theoretical approach is given. Then, our benchmark set is introduced, followed by a description of the computational details for each code. We discuss our results by starting with the KS-DFT data and then moving to the $G_0W_0$ results.

\section{Theoretical background: a summary}

The $G_0W_0$ approach has been described in numerous publications, including 
Refs.~\cite{martin2016interacting,reviewgw2019,Onida2002}, 
to which we refer for a comprehensive description. We only provide a summary in this section.

The quasiparticle (QP) energy $\epsilon_{n\mathbf{k}}^{\textrm{QP}}$ of a state with band index $n$ and wavevector $\mathbf{k}$ is calculated from the linearized self-energy equation
\begin{equation}
    \epsilon_{n\mathbf{k}}^{\textrm{QP}} = \epsilon_{n \mathbf{k}}^{\textrm{KS}} + Z_{n \mathbf{k}} \langle \psi_{n\mathbf{k}} | \Sigma^{GW}(\epsilon_{n \mathbf{k}}^{\textrm{KS}}) - v_{\textrm{xc}}^{\textrm{KS}} | 
    \psi_{n\mathbf{k}} \rangle ,
    \label{QPE}
\end{equation}
where $\epsilon_{n \mathbf{k}}^{\textrm{KS}}$ and $\psi_{n\mathbf{k}}$ are the corresponding KS eigenenergy and wavefunction, obtained from a prior DFT calculation. The renormalization factor $Z_{n \mathbf{k}}$ is defined as
\begin{equation}
    Z_{n\mathbf{k}} = \left[1 - \left.\frac{d}{d\omega} \langle \psi_{n\mathbf{k}} | \Sigma^{GW}(\omega) | \psi_{n\mathbf{k}} \rangle \right|_{\omega=\epsilon_{n \mathbf{k}}^{\textrm{KS}}}  \right]^{-1} .
\end{equation}
The self-energy $\Sigma^{GW}(\omega)$ is a two-point function (arguments are not made explicit for the sake of simplicity). In the $GW$ approach, it is computed from~\cite{HedinGW}
\begin{eqnarray}
    \Sigma^{GW}(\omega) = \frac{\text{i}}{2\pi} \int 
    d\omega' G(\omega+\omega') W(\omega') e^{\text{i}
    \omega'\eta} ,
    \label{SigmaGW}
\end{eqnarray}
where $G$ is the single-particle Green function (a two-point function), obtained from KS wavefunctions and eigenenergies. $W$ is the screened Coulomb potential (also a two-point function) treated at the level of the random phase approximation (RPA),
\begin{eqnarray}
    W(\omega) = \varepsilon^{-1}(\omega)v,
\end{eqnarray}
with $\varepsilon^{-1}$ being the inverse of the microscopic dielectric function $\varepsilon$, 
\begin{eqnarray}
    \varepsilon(\omega) = 1-v\chi_0(\omega),
    \label{MicroDielFunc}
\end{eqnarray}
$\chi_0$ the independent-particle polarizability, and $v$ the bare Coulomb potential. 
The polarizability $\chi_0$ is computed from the Adler-Wiser approach, also based on KS wavefunctions and eigenenergies~\cite{Hybertsen1986}.

These equations are valid for all codes involved here. However, each of them is based on different approaches in terms of the basis set used to expand the KS wavefunctions, the potential, and the two-point functions. The latter also lead to differences in the handling of the $GW$ equations. The particular details for each code will be presented in Section~\ref{sec:computational_details}.

\section{\label{sec:materials}Materials}
We have chosen a set of six materials, \ie \ce{Si} (diamond structure), \ce{TiO2} (rutile structure), \ce{ZnO} (wurtzite structure), \ce{ZrO2} (cubic structure), \ce{Zr2Y2O7}, and \ce{MoWS4}, for our comparison of $G_0W_0$ calculations performed with four codes (Abinit, \exciting, FHI-aims, and GPAW). 

The first three materials were selected as they are frequently discussed in the literature. Si is a standard benchmark system~\cite{rangel2020reproducibility} when developing new electronic-structure methods, including a variety of $GW$ flavors\cite{bruneval2006effect}. \ce{ZnO} and \ce{TiO2}, in turn, present numerical challenges, as previous $G_0W_0$ studies have shown deviations in band gaps, scattering in the range of $\sim$2~eV (see Ref.~\cite{rangel2020reproducibility} for further discussion). The reasons for such deviations are mainly (i) different approaches to frequency convolutions (different plasmon-pole models versus full frequency integration), (ii) slow convergence with respect to the number of empty bands and the representation of the dielectric screening, and (iii) the use of pseudopotentials of different applicability to the $GW$ problem. 

\ce{ZrO2} and \ce{Zr2Y2O7} are small-scale prototypes (3 and 11 atoms respectively) of yttria-stabilized zirconia (YSZ), a material used in a wide range of applications due to its excellent mechanical and thermal properties, such as solid oxide fuel cells (SOFCs), thermal barrier coatings, and dental and orthopedic implants. A realistic model of YSZ consists of at least 90 atoms, and a detailed $GW$ study would present a challenge. 

\ce{MoWS4} is a hypothetical bulk layered material with alternating \ce{MoS2} and \ce{WS2} layers. It is related to a twisted bilayer of \ce{MoS2} and \ce{WS2}, an interesting two-dimensional system with a large number of atoms due to the size of the two-dimensional primitive cell. In the present study, we focus on the numerical verification, so we consider simply a reduced bulk system with six atoms per unit cell, inspired by the existing bilayer configuration.

Key structural information is given in Table~\ref{tab.structures} (see the Supplemental Material~\cite{supp} for the atomic positions).

\begin{table*}[ht]
  \begin{center}
    \caption{Formula unit of the systems, number of atoms in the primitive cell, Bravais lattice and space group, crystallographic parameters, and wavevector sampling of the Brillouin zone for the six systems used for the current benchmark.}
    \label{tab.structures}
    \begin{tabular}{lcllrrrrrrrr}
    \hline
      
{System} & \textbf{N} & {Bravais lattice} &{Space group} & a (\AA) & b (\AA) & c (\AA) & $\alpha$ & $\beta$ & $\gamma$ & k-point grids \\
       \hline
      \ce{ZrO2} & 3  &  Face-centered cubic (fcc)  & $Fm\bar{3}m$ & 3.586 & 3.586& 3.586 & 60& 60 & 60 & 6$\times$6$\times$6\\
      \ce{Zr2Y2O7} & 11 &  Triclinic & $P1$& 5.171 & 5.252  & 5.373 & 90  & 90 & 90 & 4$\times$4$\times$4 \\
      \ce{MoWS4} & 6 & Hexagonal &$P3 m 1$&3.148 & 3.148  & 12.418 & 90  & 90 & 120& 6$\times$6$\times$2\\
      \ce{Si} & 2 &  Face-centered cubic (fcc) & $Fd\bar{3}m$& 3.839 & 3.839  & 3.839 & 60  & 60 & 60& 8$\times$8$\times$8\\
      \ce{TiO2} & 6 &  Primitive tetragonal & $P4_2/mnm$&4.595 & 4.595  & 3.128 & 90  & 90 & 90&6$\times$6$\times$10\\
    \ce{ZnO} & 4 &  Hexagonal & $P6_3mc$& 3.249  & 3.249 & 5.204 & 90  & 90 & 120& 8$\times$8$\times$5\\
      \hline
    \end{tabular}
  \end{center}
\end{table*}

\section{\label{sec:computational_details}Computational details}

We use the Perdew–Burke–Ernzerhof exchange correlation functional~\cite{PBE_orig,PBE_orig_erratum} optimized for solids (PBEsol)~\cite{PBE_sol,PBE_sol_erratum} as the DFT starting point for $G_0W_0$. To maintain consistency across calculations, we have standardized the Brillouin zone sampling for all systems for all codes. This uniform approach aids in fair comparisons between different code implementations. These grids are also reported in Table~\ref{tab.structures}.

For the band gaps (KS or QP levels), we  focus on high-symmetry points. 
The (in)direct band gap is not always located at wavevectors set by symmetry.
For silicon, the valence band maximum is at $\Gamma$, while the conduction band minimum lies along the $\Gamma-$X line, closer to X than to $\Gamma$. Precisely pinpointing the conduction band minimum is not essential for our comparison, as its location may vary slightly between codes. Therefore, we examine the lowest conduction band eigenenergy at $\Delta=(3/8,3/8,0)$, approximating the fundamental gap as that between $\Gamma$ and $\Delta$.
For the \ce{MoWS4} bulk system, we take a similar approach: the conduction band minimum is approximated at the high-symmetry point $\frac{1}{2}$K$\Gamma=(1/6,1/6,0)$, while the valence band maximum remains at $\Gamma$. 
Thus, the fundamental gap is roughly estimated as the separation between K and $\frac{1}{2}$K$\Gamma$.
Fixing the wavevectors improves the consistency of gap values across all codes.

\subsection{\label{sec:abinit}Abinit}
Abinit is a planewave-based code, in which the core electrons are taken into account thanks to either norm-conserving pseudopotentials~\cite{Hamann2013,kleinman1982efficacious} or projector augmented-waves (PAW)~\cite{blochl1994projector,torrent2008implementation,kresse1999ultrasoft}. The $GW$ methodology is considered adequate only for regular norm-conserving pseudopotentials or for a special kind of PAW atomic dataset, in which the norm-conservation of partial waves is retained~\cite{klimevs2014predictive}. Hence, in the present study, we employ norm-conserving pseudopotentials from the PseudoDojo project~\cite{Setten2018} with the following core-valence partitioning: \ce{Zr} (12 valence electrons), \ce{O} (6 valence electrons), \ce{W} (14 valence electrons), \ce{S} (6 valence electrons), \ce{Mo} (14 valence electrons), \ce{Si} (4 valence electrons), \ce{Ti} (12 valence electrons), and \ce{Zn} (20 valence electrons). This selection ensures a precise representation of atomic interactions within our system.\\ 

We have fine-tuned our underlying DFT calculations by choosing the planewave energy cutoff ($E_{\text{cut}}$) to guarantee convergence of total energies within 20 $\mathrm{meV}$ per atom. For the ${G_0W_0}$ calculations, the relevant parameters (including energy cutoff for the susceptibility and the dielectric (screening) matrix ($E_\textrm{c}^{\chi}$), and the number of unoccupied bands to generate the susceptibility, the screening and the self-energy operator) are tuned to achieve convergence of QP energies within a narrow margin of 20 $\mathrm{meV}$. This careful calibration ensures that the calculated QP energies are highly accurate and reliable for the system under investigation. The optimized values of the relevant parameters for the ground state and $G_0W_0$ calculations are given in Table~\ref{tab.abinit_parameter}. Note that nband is the total number of bands considered (occupied and unoccupied) in the calculation. 

Except for \ce{Zr2Y2O7}, we evaluate the self-energy along the imaginary frequency axis using a linear mesh extending up to 50 eV and 60 points, then use analytic continuation to obtain it for real frequencies, and calculate the QP energies from the linearized QP equation, Eq.(\ref{QPE}). For the \ce{Zr2Y2O7} system, due to an observed instability of the analytic continuation approach, we employ the contour deformation (CD) method, utilizing 60 frequency points along the imaginary axis extending up to 10 eV, and 100 points along the real axis to achieve convergence criterion. 

\begin{table}[ht]
  \begin{center}
    \caption{Numerical parameters for the selected systems treated with Abinit.}
    \label{tab.abinit_parameter}
    \begin{tabular}{lccrr} 
    \hline      
{System} & $E_{\text{cut}}$(Ha)* & $E_\textrm{c}^{\chi}$(Ha) &nband \\
       \hline
      \ce{ZrO2} & 40  & 8  & 350  \\
      \ce{Zr2Y2O7} & 35 & \{2, 4, 6, 8\}& 2000 \\      
      \ce{MoWS4} & 40  & 6 &1000 \\
      \ce{Si} & 20 & 10 & 500  \\
      \ce{TiO2} & 40 & \{2, 4, 6, 8, 10\} & 2000 \\      
      \ce{ZnO} & 48  & 12  & 2000 \\
      \hline
    \end{tabular}
  \end{center}
\end{table}
Achieving convergence within our target criterion (20 meV) for \ce{Zr2Y2O7} and rutile \ce{TiO2} posed a significant challenge given our existing computational resources and the standard double convergence study involving the cutoff energy in the screened interaction and the number of bands. Consequently, we needed to leverage both input variables and mathematical techniques to address this obstacle. To this end, we employed the methodology proposed by Gulans~\cite{gulans2014towards}. Utilizing the formula presented in this reference, 
\begin{eqnarray}\label{extrapolation}
\Delta E_\textrm{g}(E_\textrm{c}^{\chi}) = 
\Delta E_\textrm{g} (\infty) 
&+& B_3 \times (E_c^{\chi})^{-3/2}\nonumber\\
&+& B_5 \times (E_c^{\chi})^{-5/2},   
\end{eqnarray}
one could exploit the extrapolation technique to obtain the band gap of the structure. In Eq.~(\ref{extrapolation}), $\Delta E_\textrm{g}$ represents the difference between QP and KS band gap, $E_\textrm{c}^{\chi}$ denotes the cut-off kinetic energy for the dielectric matrix in the planewave basis, $B_3$ and $B_5$ are the fitting parameters and $\Delta E_g (\infty)$ represents the asymptotic (converged) $GW$ energy of the system, obtained thanks to the extrapolation procedure. The corresponding results are shown in Table~\ref{c1} and Table~\ref{tio2}. 

\begin{table}[ht]
        \footnotesize
        \caption{Fitting values for $\mathrm{Zr_2Y_2O_7}$}
        \centering
  \begin{tabular}{cclrrr}
    \hline
    Set of $E_\textrm{c}^{\chi}$ (Ha) & nband & HSP& $B_3$ & $B_5$ &
    $\Delta E_{\textrm{g}} (\infty)$ \\
    \hline
     $\{2,4,6,8\}$ & 2000& Y$-$Y& -2.799 &  3.633 & 1.510\\
     $\{2,4,6,8\}$ & 2000& $\Gamma-\Gamma$& -3.369 &  4.100& 1.420\\
     $\{2,4,6,8\}$ & 2000& X$-$X& -2.715 &  3.629 & 1.528\\
     \hline
\end{tabular}
\label{c1}
\end{table}
\begin{table}[ht]
        \footnotesize
        \caption{Fitting values for rutile $\mathrm{TiO_2}$}
        \centering
        \begin{tabular}{ccccc} 
        \hline
        Set of $E_\textrm{c}^{\chi}$ (Ha)& nband & $B_3$ & $B_5$ & $\Delta E_{\textrm{g}} (\infty)$\\
        \hline
$\{2,4,6,8,10\}$ & 2000 & -2.625 &   6.089  &     1.555\\
\hline
\end{tabular}
\label{tio2}
\end{table}
All inputs and outputs of the calculations carried out using Abinit are available in an open-data repository\cite{Abinit_data}.

\subsection{\label{sec:exciting}\exciting{}}

\exciting{} is based on the LAPW+lo method, where the basis consists of augmented planewaves and local orbitals~\cite{Slater1937,Sjostedt2000}. Previous $G_0W_0$ studies using this basis showed that the QP energies are not sensitive to the choice of the APW cutoff as long as the KS gaps are converged~\cite{friedrich2011band,Nabok2016}. However, it is important to employ extensive sets of local orbitals in $GW$ calculations, and these sets go far beyond what is needed for a precise ground-state calculation. We add these additional local orbitals to $\ell$-channels up to values of 5--6, depending on the atomic species, and construct them from the solutions of the radial Schrödinger equation with correspondingly high energy parameters. Such basis functions are known in the literature as high-energy local orbitals (HELO)~\cite{Michalicek2013}. To preserve their linear independence in a sequence of HELOs with the same $\ell$, we ensure that each of them has one additional radial node compared to the next one lower in energy. Sets of HELOs are constructed by adding as many of them as needed to each $\ell$-channel until the changes in the QP gap are below 1--2 meV.

The \exciting{} $G_0 W_0$ implementation is formulated in terms of an auxiliary mixed-product basis that is used to express non-local quantities such as the dielectric tensor and the screened Coulomb potential~\cite{jiang2013fhi}. The QP self-energies are computed on a non-uniform grid of 32 imaginary frequencies; denser grids have a negligible effect on the QP energies for the systems studied. We evaluate the self-energy at real frequencies via analytic continuation and calculate the QP energies from the linearized QP equation, Eq.(\ref{QPE}). Since the QP energies converge slowly with the number of empty bands, an extrapolation technique specific to LAPW+lo is employed. This makes it possible to obtain results corresponding to the complete-basis limit, but there remains a numerical uncertainty stemming from the considered range of empty bands. For this reason, we use all available empty bands despite the very high cost of such an approach, especially for materials with large unit cells. In particular, the required effort for the \ce{Zr2Y2O7} calculation with the chosen Brillouin zone sampling is excessive. We address this problem by employing the following estimate: 

\begin{align}
    &E_\mathrm{g}(\mathrm{fine~}\mathbf{k},\mathrm{all~bands}) \approx E_\mathrm{g}(\mathrm{fine~}\mathbf{k},\mathrm{few~bands}) + \nonumber\\
    &\quad E_\mathrm{g}(\mathrm{coarse~}\mathbf{k},\mathrm{all~bands}) - E_\mathrm{g}(\mathrm{coarse~}\mathbf{k},\mathrm{few~bands}),
\end{align}

where fine and coarse $\mathbf{k}$-grids are $4\times 4\times 4$ and $2\times 2\times 2$, respectively. Using this approach, we find that the  \ce{TiO2} band gap is already converged within the target precision at the $4\times 4\times 6$ $\mathbf{k}$-point mesh. 

All input and output files of the calculations carried out using \exciting{} are available in NOMAD~\cite{Scheidgen2023}, DOI: \cite{exciting_data} .

\subsection{\label{sec:fhi-aims} FHI-aims} 

FHI-aims is an all-electron electronic structure code, which is based  on numeric atom-centered orbitals (NAOs)~\cite{FHIaims}. To expand the KS orbitals we employ the hierarchical \textit{tier 2} NAO basis functions~\cite{FHIaims}. Similarly as for the LAPW case, we supplement our basis set for $GW$ calculations with highly localized orbitals, specifically, Slater-type orbitals (STOs) up to $\ell=7$. Four even-tempered STOs are included in each angular momentum channel, see Table~\ref{fhiaims_basis_param} for details. We incrementally added STOs with higher $\ell$ values until achieving convergence of the QP gap. Including STOs up to $\ell=5$ achieves the convergence of the QP gap within 50 meV (see Figures S1-S4 in the supporting information (SI)). Further extension of the STOs to $\ell=6$ improves the convergence of the QP gap to within 10 meV (see Figures S1 and S2 in the SI). All unoccupied bands resolved by our basis set are included in the evaluation of polarizability and self-energy.
\begin{table}[ht]
    \centering
    \caption{Basis sets used for FHI-aims calculations. Primary basis functions and extra ABFs added to the autogenerated set.}
    \label{fhiaims_basis_param}
\begin{tabular}{lllll}
\hline
System & NAO & &  STO & Extra functions for ABF set \\ \hline
    \ce{ZrO2} & \textit{tier 2} & & s,p,d,f,g,h & 4f,5g \\ 
    \ce{Zr2Y2O7}& \textit{tier 2} & & s,p,d,f & 4f,5g \\ 
    \ce{MoWS4} & \textit{tier 2} & & s,p,d,f,g,h & 4f,5g,6h \\ 
    \ce{Si} & \textit{tier 2} & & s,p,d,f,g,h & 4f,5g, \\  
    \ce{TiO2} &  \textit{tier 2} & & s,p,d,f,g & 4f,5g    \\  
    \ce{ZnO} & \textit{tier 2} & & s,p,d,f,g,h,i & 4f,5g  \\ \hline 
\end{tabular}

\end{table}

In FHI-aims, the self-energy $\Sigma_{n}(\textit{i}\omega)$ is reformulated using the resolution-of-the-identity (RI) approach ~\cite{RI_Whitten73,RI_Dunlap79,RI_Mintmire82,RI_Vahtras93,ren2012resolution}, which transforms the four-center two-electron Coulomb repulsion integral into three- and two-center Coulomb integrals. Our RI-based $GW$ implementation employs a localized version of RI, known as RI-LVL, which notably decreases the computational cost associated with computing and storing these integrals~\cite{Ihrig_2015,Ren/etal:2021}. The RI approach relies on auxiliary basis functions (ABFs), which are also NAOs and which are autogenerated in FHI-aims as described in Ref.~\cite{ren2012resolution}. To mitigate errors arising from the RI-LVL approximation, we included extra hydrogen-like functions on top of the one-electron basis set when generating ABFs~\cite{Ihrig_2015}. Except for MoWS$_4$, including $4f$ and $5g$ functions is sufficient and further addition of $6h$ will change the fundamental gap by less than 5 meV. For \ce{MoWS4}, however, adding 6$h$ function is necessary to avoid a pathological dielectric matrix under current treatment of Coulomb divergence (see next paragraph) for a reasonable quasi-particle band structure.

The divergence of the Coulomb interaction at the long-wavelength limit ($\mathbf{q}\to0$) is handled in the evaluation of the dielectric matrix and self-energy~\cite{Ren/etal:2021}. In FHI-aims, the dielectric matrix given Eq.\eqref{MicroDielFunc} is expanded in ABFs and is first computed using the singularity-lifted bare Coulomb matrix. Subsequently, $\varepsilon(\mathbf{q}=0)$ is corrected by the macroscopic dielectric function based on the $\mathbf{k}\cdot\mathbf{p}$ theory, using the real-space momentum matrix elements where the corresponding Hamiltonian element is larger than certain threshold ($10^{-13}$ Ha for this work). The divergence issue in $W$ for the Brillouin Zone integral of the self-energy is circumvented by adopting a truncated Coulomb operator based on the work by Spencer and Alavi~\cite{spencer2008}.

To obtain QP energies, we perform an analytic continuation of the self-energy from the imaginary to the real frequency axis  using  a modified Gauss-Legendre~\cite{ren2012resolution}. We employ 60 frequency points and a Pad\'e model~\cite{Pade_paper} with 16 parameters. All inputs and outputs of the calculations carried out using FHI-aims are available in an open-data repository~\cite{fhiaims_data} 

\subsection{\label{sec:gpaw}GPAW}
GPAW is an open source Python package for electronic structure calculations~\cite{mortensen2024gpaw,enkovaara2010electronic} based on the projector augmented wave (PAW) method~\cite{blochl1994projector}. A unique feature of GPAW is that it supports three different wavefunctions representations, namely real space grids, numerical atomic orbitals, and planewaves. The $GW$ method is implemented in GPAW for the planewave representation only~\cite{huser2013quasiparticle,huser2013dielectric}. 
The QP energies are obtained by evaluating the diagonal matrix elements of the QP equation, Eq.~(\ref{QPE}), with the self-energy obtained using full frequency integration along the real frequency axis for evaluating the convolution of $G$ and $W$, Eq. (\ref{SigmaGW}). All dynamical quantities, including the spectral function, can be obtained as a function of frequency without relying on analytical continuation. 

Custom norm-conserving (NC) PBEsol setups were generated specifically for the calculations, and in some cases the cutoff radii were shortened compared to the normal soft PAW setups. Projector functions were added initially for each valence state, then new ones added until very good logarithmic derivative over the angular momentum channels up to 10 Hartrees was obtained. In Table~\ref{gpaw_setups}, we list the most important parameters used to generate the PAW setups. They are used as input for GPAW's setup generator. The second column contains the cutoff radius of the PAW projectors. The third and fourth columns contains projectors for each of the angular channels that are included.
For Mo, W, S, Ti, Zn, and O, the local potential was also fitted to produce correct scattering at 0 Hartree in the f-channel.

GPAW performs automatic extrapolation of results, by calculating the QP energy shifts with several cutoffs simultaneously. For each energy cutoff, the irreducible response matrix is built with the highest 
cutoff but only up to unoccupied bands up to the specific cutoff. Then, the dielectric matrix is truncated to the cutoff at hand to get $W$, and self-energy matrix elements are evaluated with $G$ bands going also up to the specific cutoff. This way, one obtains three values for each QP energy shift, which are subsequently fitted using the polynomial E = $E_{\infty} + \alpha E_{c}^{-3/2}$. This way, one obtains simultaneous extrapolation of effects of $G$ band cutoff, irreducible response matrix bands cutoff, and dielectric matrix planewave cutoff with minimal efforts. All values reported are extrapolated automatically this way. The extrapolation cutoffs are chosen to be on the very high end of the cutoffs, for example 18, 19 and 20 Hartree. This way, one assures to be sufficiently in the asymptotic regime of convergence. Indeed, the minimum $R^2$ coefficients of the polynomial fits are typically larger than 0.99. All calculations can be found at \cite{GPAWdata}.

In Table \ref{gpawparameters}, the planewave cutoffs for the DFT ground state and the dielectric matrix as well as the number of bands included for the latter are provided (the maximum cutoffs used for extrapolation). The number of bands for $G$ and $\chi_0$ are automatically selected from the cutoff. It is an estimate of how many reciprocal space points the sphere with radius of a Fermi wavevector of the cutoff covers. They are only displayed for comparison to the other codes.

\begin{table}[ht]
\caption{GPAW parameters for custom NC PAW setups: cut-off radii (in Bohr) for s,p,d channels, projector parameters for valence and for extra projectors. The valence projector functions are specified by atomic $nl$ quantum numbers. For extra projectors the notation is $El$, where $E$ is the energy (in Ha) of the (unbound) state. }

\begin{tabular}{llll}
\hline
Element & Radii (s,p,d) & Valence projectors  & Extra projectors \\
\hline
\ce{Zr}      & 2.0,2.0,2.0   & 4s,5s,4p,5p,4d & 5.0s,5.0p,2.5d,8.0d,3.0f \\
\ce{Y}       & 2.0,2.0,2.0   & 4s,5s,4p,5p,4d & 5.0s,5.0p,2.5d,8.0d,3.0f \\
\ce{Mo}      & 2.3,2.3,2.3   & 4s,5s,4p,5p,4d & 0.86d \\
\ce{W}       & 2.3,2.3,2.0 & 5s,6s,5p,6p,5d & 0.78d \\
\ce{S}       & 1.3,1.3,1.3   & 3s,3p & 8.0s,8.0p,0.0d,8.0d \\
\ce{Ti}      & 2.1,2.1,2.1   & 3s,4s,3p,4p,3d & 4.0s,4.0p,1.0d,7.0d  \\
\ce{Zn}      & 2.1,2.1,2.1   & 3s,4s,3p,4p,3d & 4.0s,4.0p,1.0d,7.0d \\
\ce{O}       & 1.1,1.1,1.1   & 2s,2p & 4.0s,4.0p,0.0d \\
\hline
\end{tabular}
\label{gpaw_setups}
\end{table}

\begin{table}[h]
\caption{GPAW ground state cut-off energy ($E_{\rm c}^{\rm gs}$), dielectric matrix cut-off energy $(E_{\rm c}^{\rm diel})$, and number of bands, nband.}
\begin{tabular}{lrrrr}
\hline
System & $E_{\rm c}^{\rm gs}$ (Ha) & $E_{\rm c}^{\rm diel}$ (Ha) & nband\\
\hline
\ce{ZrO2}     & 55      & 20                  & 940                 \\
\ce{Zr2Y2O7}  & 44      & 10                  & 1487                \\
\ce{MoWS4}    & 58      & 10                  & 1086                \\
\ce{Si}       & 44      & 10                  & 411                 \\
\ce{TiO2}     & 55      & 30                  & 3306                \\
\ce{ZnO}      & 55      & 20                  & 1371                \\
\hline
\end{tabular}
\label{gpawparameters}
\end{table}

\section{\label{sec:result}Results and discussion}

\subsection{Kohn-Sham band gaps }

\begin{table}[h]
    \centering
    \caption{PBEsol band gaps (direct and indirect) in eV across the indicated high-symmetric points (HSP) obtained from the four community codes used in this work.} 
    \label{tab:PBEsolbandgap}
\begin{tabular}{llcccc}
\hline
System	&	HSP	&	Abinit	&	\exciting{}	&	FHI-aims	&	GPAW \\ \hline
    \ce{ZrO2} 	&	 $\Gamma-\Gamma$ 	&	3.867	&	3.863	&	3.854	& 3.847	\\
    \ce{ZrO2} 	&	 X$-\Gamma$ 	& 3.324	&	3.318	&	3.309	&	3.293 	\\
    \ce{ZrO2} 	&	 X$-$X 	&	3.748	&	3.748	&	3.741	&	3.733	\\ \hline
    \ce{Zr2Y2O7} 	&	 Y$-$Y 	&	3.374  	&	3.374	&	3.362	&	3.373	\\
    \ce{Zr2Y2O7} 	&	 Y$-\Gamma$ 	&	2.760	&	2.756	&	2.742	&	2.747	\\
    \ce{Zr2Y2O7} 	&	 $\Gamma-\Gamma$ 	&	2.817	&	2.813	&	2.798	&	2.798	\\
    \ce{Zr2Y2O7} 	&	 X$-\Gamma$ 	&	2.817 	&	2.810	&	2.798	&	2.759 \\
    \ce{Zr2Y2O7} 	&	 X$-$X 	&	3.279	&	3.279	&	3.267	&	3.269	\\ \hline
    \ce{MoWS4} 	&	 K$-$K 	&	  	1.549 &	1.549	&	1.551	&	1.539	\\
    \ce{MoWS4} 	&	 $\Gamma-\frac{1}{2}$K$\Gamma$ 	&	0.922	&	0.923	&	0.923	&	0.921	\\
    \ce{MoWS4} 	&	 $\Gamma-\Gamma$ 	&	2.118  	&	2.119	&	2.118	&	2.116	\\ \hline
    Si 	&	 $\Gamma-\Gamma$ 	&	2.513	&	2.522	&	2.515	&	2.520	\\
    \ce{Si} 	&	 $\Gamma-\Delta$ 	&	0.506	&	0.510	&	0.506	&	0.508	\\
    \ce{Si} 	&	 $\Gamma-$X 	&	0.593	&	0.596	&	0.596	&	0.594	\\
    \ce{Si}	&	 X$-$X 	&	3.471	&	3.470	&	3.479	&	3.469	\\ \hline
    \ce{TiO2} 	&	 $\Gamma-\Gamma$ 	&	1.834	&	1.809	&	1.794	&	1.733 \\ \hline
    \ce{ZnO} 	&	 $\Gamma-\Gamma$ 	&	0.762	&	0.714	&	0.713	&	0.701 \\ \hline
\end{tabular}
\end{table}

Table~\ref{tab:PBEsolbandgap} presents a comparison of the calculated PBEsol band gaps (in eV) at selected high-symmetry points for the selected materials across four codes used in this study. For \ce{ZrO2}, the values are very close to each other, with the largest, though still a small difference is found for the $X-\Gamma$ gap, where GPAW and Abinit differ by 31 meV. The differences between the all-electron codes are even smaller, remaining within 10 meV.

In the case of \ce{Zr2Y2O7}, there is slightly more variation, with FHI-aims generally producing lower band-gap values compared to Abinit and \exciting{}, particularly across Y$-\Gamma$ and $\Gamma-\Gamma$. The largest discrepancy occurs for the $X-\Gamma$ gap, with a 58 meV difference between GPAW and Abinit, while the all-electron codes differ by no more than 15 meV. For \ce{MoWS4}, the direct band gap at the K$-$K point is relatively consistent across all codes, with only a slightly higher value seen in FHI-aims (1.551 eV) compared to GPAW (1.539 eV), a difference of 12 meV. This represents the largest discrepancy for \ce{MoWS4}.

In the case of silicon (\ce{Si}), the results show minimal variation, particularly at the $\Gamma$ point, where the band gaps range between 2.513 eV (Abinit) and 2.522 eV (\exciting) only. The largest difference occurs between FHI-aims and GPAW for the X$-$X gap. 

For \ce{TiO2} and \ce{ZnO}, GPAW generally produces lower band gaps than the other codes. The largest discrepancy of 101 meV for \ce{TiO2}, occurs between Abinit and GPAW, while the differences among the all-electron codes remain within 15 meV. Also for \ce{ZnO}, the largest discrepancy of 61 meV is between Abinit and GPAW, while there is just a 1 meV difference among the all-electron codes. 

Overall, the largest discrepancy among all considered materials across all four codes is 101 meV, while the maximum discrepancy between the all-electron codes is only 15 meV. The error of the pseudopotential codes is attributed to the limitations in describing the core states of the oxygen atom.

\begin{table*}[ht]
    \centering
    \caption{Median, mean absolute deviation (MAD), and maximum absolute deviation (maxAD) in meV analyzed for the four codes. The top table presents values for the PBEsol functional, while the bottom table shows the same for $G_0W_0$@PBEsol.} \label{tab:deviations}
    \begin{tabular}{lrrrlrrrlrrrlrrr}
\multicolumn{15}{c}{PBEsol}\\ \hline		
	&	\multicolumn{3}{c}{Abinit} &	& \multicolumn{3}{c}{\exciting{}}	&	&	\multicolumn{3}{c}{FHI-aims} &	&	\multicolumn{3}{c}{GPAW}	\\ \cline{2-4} \cline{6-8} \cline{10-12} \cline{14-16}
	&	Median	&	MAD	&	maxAD	&	&	Median	&	MAD	&	maxAD	&	&	Median	&	MAD	&	maxAD	&	&	Median	&	MAD	&	maxAD	\\ \cline{2-4} \cline{6-8} \cline{10-12} \cline{14-16}
Abinit	&	- -	&	- -	&	- -	&	&	0	&	7	&	48	&	&	12	&	13	&	49	&	&	10	&	21	&	101	\\
\exciting	&	0	&	7	&	48	&	&	- -	&	- -	&	- -	&	&	7	&	8	&	15 &	&	10	&	15	&	76	\\
FHI-aims	&	-12	&	13	&	49	&	&	-7	&	8	&	15	&	&	- -	&	- -	&	- -	&	&	2	&	12	&	61	\\
GPAW	&	-10	&	21	&	101	&	&	-10	&	15	&	76	&	&	-2	&	12	&	61	&	&	- -	&	- -	&	- -	\\ \hline \\
	\multicolumn{15}{c}{$G_0W_0$@PBEsol}\\ \hline		
	&	\multicolumn{3}{c}{Abinit} &	& \multicolumn{3}{c}{\exciting{}}	&	&	\multicolumn{3}{c}{FHI-aims} &	&	\multicolumn{3}{c}{GPAW}	\\ \cline{2-4} \cline{6-8} \cline{10-12} \cline{14-16}
&	Median	&	MAD	&	maxAD	&	&	Median	&	MAD	&	maxAD	&	&	Median	&	MAD	&	maxAD	&	&	Median	&	MAD	&	maxAD \\ \cline{2-4} \cline{6-8} \cline{10-12} \cline{14-16}
Abinit	&	- -	&	- -	&	- -	&	&	-52	&	96	&	265	&	&	9	&	82	&	213	&	&	-50	&	119	&	267	\\
\exciting	&	52	&	96	&	265	&	&	- -	&	- -	&	- -	&	&	-2	&	90	&	225	&	&	-33	&	63	&	334	\\
FHI-aims	&	-9	&	82	&	213	&	&	2	&	90	&	225	&	&	- -	&	- -	&	- -	&	&	2	&	119	&	265	\\
GPAW	&	50	&	119	&	267	&	&	33	&	63	&	334	&	&	23	&	119	&	265	&	&	- -	&	- -	&	- -	\\ \hline
    \end{tabular}
\end{table*}

In addition to the analysis of code performance for the individual materials, the data in Table~\ref{tab:PBEsolbandgap} can also be examined from a statistical perspective, though the dataset is relatively small. A choice must also be made regarding whether to assign equal weight to each material or to each entry in Table~\ref{tab:PBEsolbandgap}. Given the limited number of samples of the analysis, a simple approach was chosen. In the following, we show the median, mean absolute deviation (MAD), and maximum absolute deviation (maxAD) for each code pair, giving equal weight to each entry in the Table \ref{tab:deviations}. In the top table, the statistics for the PBEsol band gaps are show. The PBEsol results show excellent agreement (MAD below 10 meV) between Abinit and \exciting{} (7 meV) as well as between \exciting{} and FHI-aims (8 meV), both of which are all-electron codes. The largest MAD, 21 meV, is observed between the two plane-wave codes, Abinit and GPAW. 

\begin{figure*}{ht}
    \centering
    \includegraphics[width=1.0\linewidth]{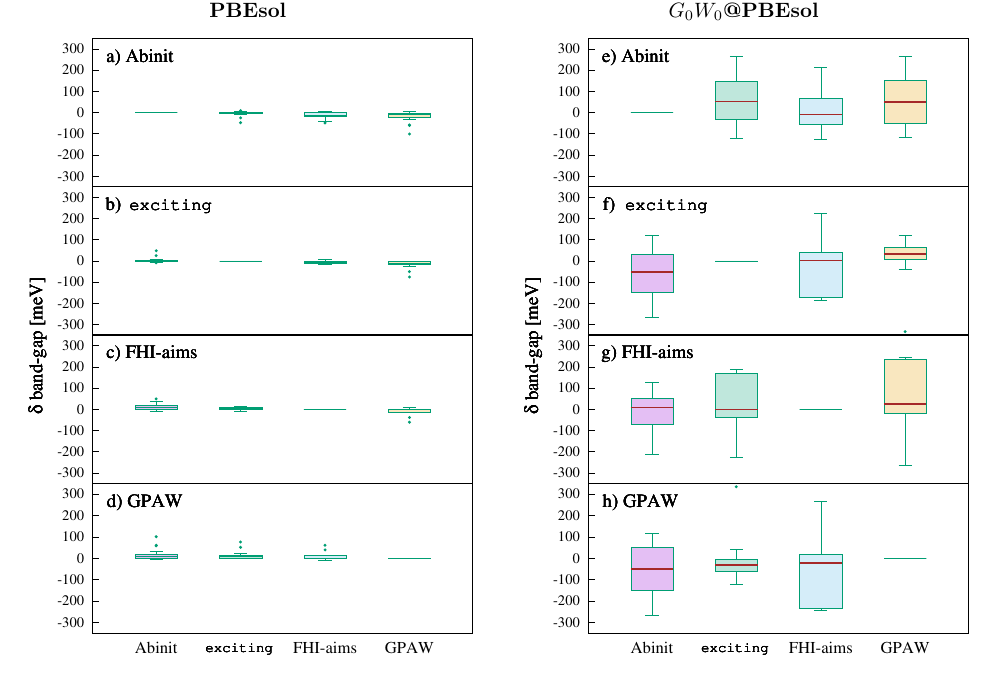}
    \caption{Pairwise differences $\delta$ of band gaps between two codes, $\delta=\text{code}_i - \text{code}_{\text{ref}}$, in meV, where the subscript $i$ specifies the code on the horizontal axis, and the subscript ``ref" denotes the code used as reference. Left panel (a-d): differences at the KS-DFT level using the PBEsol functional. Right panel (e-h): differences at the $G_0W_0@$PBEsol level of theory. The boxes represent the interquartile range, showing where most of the data is concentrated. Points lying beyond the boxes are considered outliers. The red line within each box represents the median of the distribution.}
    \label{fig:deltabandgap}
\end{figure*}

The comparison between codes is further illustrated in Fig. \ref{fig:deltabandgap}.
The left panel of Fig. \ref{fig:deltabandgap} displays the differences in PBEsol band gaps between each two codes. The errors are generally within a range of 15 meV, with only a few outliers observed among this pairwise comparisons.

\subsection{$G_0W_0$ band gaps}

\begin{table}[ht]
    \centering
    \caption{Same as Table \ref{tab:PBEsolbandgap} but for $G_0W_0$@PBEsol.}
\label{GW_bandgap}
\begin{tabular}{llcccc} \hline
System	&	HSP	& Abinit&	\exciting{}	&	FHI-aims	&	GPAW \\ \hline
    \ce{ZrO2} 	&	 $\Gamma-\Gamma$ 	&	5.914	&	5.883	&	5.935	&		5.916  \\
    \ce{ZrO2} 	&	 X$-\Gamma$ 	&	5.298	&	5.319	&	5.366	&	5.348	\\
    \ce{ZrO2} 	&	 X$-$X 	&	5.616	&	5.635	&	5.598	&	 5.615	\\ \hline
    \ce{Zr2Y2O7}  	&	 Y$-$Y 	&	4.884$^\text{a}$		&	5.051	&	4.875	&	5.114	\\
    \ce{Zr2Y2O7}  	&	 Y$-\Gamma$ 	&	4.270$^\text{a}$		&	4.352	&	4.164	&	4.407	\\
    \ce{Zr2Y2O7} 	&	 $\Gamma-\Gamma$ 	&	4.237$^\text{a}$		&	4.408	&	4.221	&	 4.465	\\
    \ce{Zr2Y2O7}  	&	 X$-\Gamma$ 	&	4.346$^\text{a}$		&	4.405	&	4.218	&	4.459	\\
    \ce{Zr2Y2O7} 	&	 X$-$X 	&	4.807$^\text{a}$	&	4.980	&	4.809	&	5.044	\\ \hline
    \ce{MoWS4} 	&	 K$-$K 	&	1.978	&	2.044	&	2.083	&	2.128	\\
    \ce{MoWS4} 	&$\Gamma-\frac{1}{2}$K$\Gamma$ 	&	1.345	&	1.397	&	1.414	&	1.484	\\
    \ce{MoWS4} 	&	 $\Gamma-\Gamma$ 	&	2.487	&	2.634	&	2.655	&	2.754	\\ \hline
    \ce{Si} 	&	 $\Gamma-\Gamma$ 	&	3.248	&	3.215	&	3.202	&	3.225	\\
    \ce{Si} 	&	 $\Gamma-\Delta$ 	&	1.257	&	1.145	&	1.147	&	 1.152	\\
    \ce{Si} 	&	 $\Gamma-$X 	&	1.366	&	1.243	&	1.252	&	1.250	\\
    \ce{Si} 	&	 X$-$X 	&	4.212	&	4.113	&	4.158	&	4.109	\\ \hline
    \ce{TiO2} 	&	 $\Gamma-\Gamma$ 	&	3.333	&	3.321	&	3.546
    & 3.281		\\ \hline
    \ce{ZnO} 	&	 $\Gamma-\Gamma$ 	&	2.613	&	2.878	&	2.758 	&	2.544	\\ 
    \hline
\end{tabular}
\caption*{$^\text{a}$ These values were computed used the contour deformation method,
see Sec.~\ref{sec:abinit}.}
\end{table}

With the differences in KS values between codes now characterized, we can proceed to the $G_0W_0$ calculations. Table~\ref{GW_bandgap} presents a comparison of the calculated $G_0W_0$ band gaps (in eV) at selected high-symmetry points for various materials across the four codes. 

For \ce{ZrO2}, the computed values at these points are very close to each other, with only minor variations. The largest discrepancy of 68~meV is observed between Abinit and FHI-aims for the X$-\Gamma$ gap. The largest difference between the two all-electron codes is 52~meV, which is only slightly smaller than the maximum discrepancy across all four codes.

For \ce{Zr2Y2O7}, larger discrepancies are observed, with GPAW consistently showing higher values, approximately 0.2 eV above those of Abinit and FHI-aims. The largest difference of 244 meV occurs for the $\Gamma-\Gamma$ gap between FHI-aims and GPAW. Among the all-electron codes, the maximum difference is also significant, reaching 187 meV.

For \ce{MoWS4}, the largest discrepancy is 267 meV for the $\Gamma-\Gamma$ gap, observed between Abinit and GPAW. Among the all-electron codes, the maximum difference is comparatively small \ie just 39~meV.

The results for \ce{Si} are more consistent than in the previous cases, though Abinit tends to produce slightly lower values. The largest discrepancy, 123~meV, occurs for the $\Gamma-X$ gap between Abinit and \exciting{}. Among the all-electron codes, the maximum difference is 45~meV.

For \ce{TiO2}, the largest difference of 265 meV is observed between FHI-aims and GPAW. Among all-electron codes, the maximum difference is 225 meV. In the case of \ce{ZnO}, the largest discrepancy is 334~meV between \exciting{} and GPAW, while the difference among the all-electron codes is 120~meV.

The same pairwise MAD analysis applied to the KS case is now performed for $G_0W_0$. The results for all six pairs of codes are presented in Table~\ref{tab:deviations}. Interestingly, the smallest MAD of 63 meV is found between \exciting{} and GPAW, followed by 82 meV between Abinit and FHI-aims. The two all-electron codes exhibit a slightly higher MAD of 90 meV. The largest MAD, 119 meV, occurs between Abinit and GPAW.

These results show a noticeable difference compared to those for the KS band gaps, also listed in Table~\ref{tab:deviations}, where \exciting{} shows the closest agreement with Abinit and FHI-aims. The MAD values for the KS band gaps range from 7 meV to 21 meV, whereas the MAD values for the $G_0W_0$ band gaps range from 63 meV to 119 meV, \ie more than five times larger on average.

Overall, the largest discrepancy across all four codes is 334 meV. This value is not much bigger than the maximum discrepancy between the all-electron codes, 225 meV.



To summarize these findings, we note that there is overall no unique trend, and the magnitude of discrepancies depends very strongly on the material. The results obtained with the all-electron codes agree more closely than for other code pairs. We explain it by the ability of systematically improving the description of the QP electronic structure in exciting and FHI-aims by adding localized basis functions thus improving the basis completeness. Nevertherless, discrepancies remain even in the all-electron case, and a more systematic analysis of the origin of these differences, potentially including more materials, requires further investigation with considerable additional computational effort. Whether adding localized functions is necessary, depends on the material, as demonstrated by our convergence studies displayed in the SI. For FHI-aims, we observe notable effects on the band gap for ZrO$_2$ (70~meV) and, to a lesser extent, for MoWS$_2$ (20-50~meV).

For cases like silicon, basis set convergence is likely achieved across all codes, as it has been shown that convergence is relatively fast~\cite{Friedrich/etal:2006}, also benefiting from error cancellation between valence and conduction bands\cite{Nabok2016}.  

To gain further insight into other sources for the deviations, it is useful to analyze the different components contributing to the larger discrepancies on the $G_0W_0$ level compared to the KS-DFT data, similar to Refs. \cite{Gomez2008} and \cite{Li2012}. These components include the exact exchange energy, the exchange-correlation energy, the correlation self-energy, and the renormalization factor, all of which contribute to the QP energy for both the conduction and valence bands. 
\begin{table}[ht]
\caption{Contributions to the quasiparticle energy for Si at the $\Gamma$ point (subscript v for the top of the valence band, and c for the lowest conduction band at $\Gamma$). All quantities are given in eV except for the renormalization factors $Z_\mathrm{v}$ and $Z_\mathrm{c}$).}
\label{tab:contributions}
    \begin{tabular}{lrrrr}\hline
                                        & Abinit	& \exciting{}	& FHI-aims	& GPAW	  \\ \hline
$\Delta\epsilon_\mathrm{KS}$	        &  2.514	&  2.522	    &  2.523  	&  2.521  \\
\hline
$\Sigma_\mathrm{v}^\mathrm{x}$	&-12.673  	&-14.456       	&-14.400   	&-14.463  \\ 
$\Sigma_\mathrm{c}^\mathrm{x}$	& -5.867  	& -7.200       	& -7.261   	& -7.229  \\ 
$\Delta\Sigma^\mathrm{x}$	            &  6.806 	&  7.256       	&  7.139   	&  7.233  \\ 
\hline
$\Sigma_\mathrm{v}^\mathrm{c}$	&0.342  	&  0.273    	&    	0.898
 & 0.222  \\ 
$\Sigma_\mathrm{c}^\mathrm{c}$	&-4.305  	& -4.350       	&    	-3.720
 & -4.361  \\ 
$\Delta\Sigma^\mathrm{c}$	            &  -4.647 	& -4.623       	&     	-4.619
 & -4.583   \\ 
 \hline
$v_\mathrm{v}^\mathrm{xc}$	    &-11.216  	&-13.305       	&-13.247   	&-13.301  \\ 
$v_\mathrm{c}^\mathrm{xc}$	    &-10.027 	&-11.577       	&-11.599   	&-11.572  \\ 
$\Delta v^\mathrm{xc}$	                &  1.189    &  1.728 	    &  1.648    &  1.729  \\ 
\hline
$\Delta\epsilon_{\textrm{QP}}$	                &  3.194	&  3.215	    &  3.188    &  3.225  \\ 
\hline
$Z_\mathrm{v}$	                &  0.755	&  0.766    	&  0.761    &  0.763  \\ 
$Z_\mathrm{c}$	                &  0.753	&  0.761	    &  0.766    &  0.758  \\ \hline
    \end{tabular}
\end{table}
Table~\ref{tab:contributions} presents these contributions to the QP energies for silicon at the $\Gamma$ point, with detailed comparisons across the four different codes. The listed quantities include key physical properties, such as the KS energy gap ($\Delta\epsilon_\mathrm{KS}$) and the QP gap ($\Delta\epsilon_{\textrm{QP}}$), as well as components of the self-energy—specifically, the exchange ($\Sigma^\mathrm{x}$) and correlation ($\Sigma^\mathrm{c}$) terms, each of which plays a crucial role in understanding the corrections to the DFT-derived energies. Additionally, the table includes renormalization factors $Z_\mathrm{v}$ and $Z_\mathrm{c}$ for the valence and conduction bands, respectively, which characterize the strength of electron-electron interactions.

The slight variations in the calculated values underscore the influence of the different methods used to compute the QP properties. For example, the exchange-correlation contributions to the self-energy ($\Delta\Sigma^\mathrm{x}$ and $\Delta\Sigma^\mathrm{c}$) vary across the codes, with the \exciting{} code yielding the most negative values, indicating stronger exchange interactions. In contrast, the overall QP gap ($\Delta\epsilon_{\textrm{QP}}$) remains consistent across the methods, with values ranging from 3.188 eV to 3.225 eV, demonstrating good agreement despite the use of different computational approaches. This further underpins the conclusions from Ref. \cite{Gomez2008} that pseudopotential codes benefit from cancellations between self-energy contributions and the corresponding xc potential.

\begin{figure}[H]
    \centering
    \includegraphics{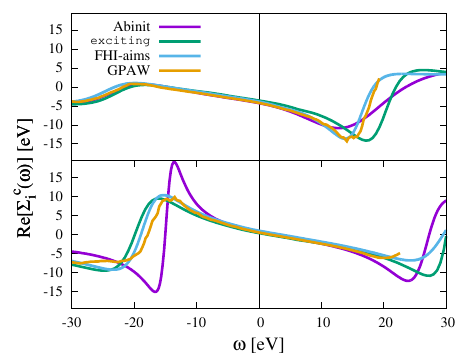}
    \caption{Real part of the correlation self-energy for bulk silicon, evaluated on the real frequency axis, at the $\Gamma$ point. The conduction band minimum is shown in the top panel, and the valence band maximum is shown in the bottom panel.}
    \label{fig:selfe_all}
\end{figure}

To better understand the discrepancies in the results, we compare the real part of the correlation self-energy for silicon as a function of energy at both the conduction band minimum (CBM) and valence band maximum (VBM) across the different codes (Fig.~\ref{fig:selfe_all}). The frequency axis is centered around the Fermi level. There is excellent agreement among all four codes within the band gap region, indicating consistency in the description of the real part of the self-energy near the Fermi level. However, significant deviations are observed for Abinit both below -10 eV and above +15 eV compared to the other three codes. These discrepancies may be attributed to the use of the norm-conserving pseudopotential approximation in Abinit.

\section{\label{sec:conc}Conclusions}
In this work, we have examined the $GW$ electronic band gaps for six selected systems with four representative codes, Abinit, \exciting{}, FHI-aims, and GPAW. The goal was to apprehend the numerical agreement/disagreement between the different codes due to different choices of basis sets and the inclusion of core electrons.

We have started the assessment at the KS-DFT level. For the KS-DFT band gaps we found a reasonable agreement between all codes, with a maximum difference of less than 0.1 eV and a mean absolute deviation of less than 21 meV. The two all-electron codes, \exciting{} and FHI-aims agree with a maximum difference of only 15 meV, and a mean absolute deviation of 8 meV.

The $G_0W_0$ results do not reach the same numerical precision as the KS-DFT results. The agreement is nevertheless on the order of 0.1-0.3~eV overall, with a maximum difference of 334 meV. The largest pairwise MAD is found to be 119 meV. The discrepancies are due to the different basis sets and the fact that \exciting{} and FHI-aims include explicitly core electrons, while Abinit and GPAW uses pseudopotentials, in which core electrons are frozen. 

Interestingly, there is a rather good agreement on the $G_0W_0$ level between \exciting{} and GPAW, being on the order of 0.1 eV, with two exceptions, the biggest being 334 meV. The MAD between \exciting{} and FHI-aims, being 90 meV, is not much smaller than the maximum MAD value of 119 meV. Values on the order of 0.1~eV give a meaningful reference for the target precision to be achieved with each code. The parameters should be tuned to reach an internal precision that should be only a small fraction of this value. This might prove challenging.

Interestingly, despite relying on different technical implementations, the numerical agreement between the all-electron codes is better than the known limitations of the $G_0W_0$ approximation itself, estimated to be about 0.5 eV. The basis set choice and treatment of core electron seem to affect the numerical precision only slightly. Still, such agreement is not fully satisfactory. Our study calls for further analyses and possible improvements of implementations, in order to obtain more predictive power of the $G_0W_0$ methodology, offering confidence in its reliability for computing electronic properties of materials.

\section{\label{sec:ack}Acknowledgment}
 This work was supported by the European Union’s Horizon 2020 research and innovation program under the grant agreement N° 951786 (NOMAD CoE). M.A. and X.G. thank Joao Abreu for information about the band gap of ZrO$_2$. This work was done in the framework of the Shapeable 2D magnetoelectronics by design project (SHAPEme, EOS Project No. 560400077525) that has received funding from the FWO and FRS-FNRS under the Belgian Excellence of Science (EOS) program. D.G. acknowledges funding from the Emmy Noether Program of the German Research Foundation (Project No. 453275048). D.G. and F.A.D acknowledge funding from the European High-Performance Computing Joint Undertaking (JU) under grant agreement No. 101118139.  K. S. T. is a Villum Investigator supported by VILLUM FONDEN (grant no. 37789). D.Z. and A.G. acknowledge the EuroHPC Joint Undertaking for awarding access to the EuroHPC supercomputer LUMI, hosted by CSC (Finland) and the LUMI consortium through a EuroHPC Development Access call. Computational resources have been provided by the supercomputing facilities of the Université catholique de Louvain (CISM/UCL) and the Consortium des Equipements de Calcul Intensif en Fédération Wallonie Bruxelles (CECI) funded by the FRS-FNRS under Grant No. 2.5020.11. The authors wish to acknowledge CSC – IT Center for Science, Finland, for computational resources. The authors gratefully acknowledge the computing time provided to them on the high-performance computer Noctua 2 at the NHR Center PC2. These are funded by the Federal Ministry of Education and Research and the state governments participating on the basis of the resolutions of the GWK for the national high performance computing ar universities (www.nhr-verein.de/unsere-partner) M.-Y. Z. acknowledges the computer resources on the high-performance computer system Raven at the Max Planck Computing and Data Facility.

\medskip
\pagebreak
\bibliography{refs.bib} 
\end{document}